\documentclass[aip, reprint]{revtex4-2}

\usepackage{graphicx} 
\graphicspath{{images/}}
\usepackage{dcolumn}
\usepackage{bm}
\usepackage{booktabs}
\usepackage{textcomp}
\usepackage{gensymb}
\usepackage{diagbox} 
\usepackage{float}
\usepackage{amsmath}
\usepackage{amssymb}

\begin{document}

\title{Reconstruction and Optimization of Coherent Synthesis by Fourier Optics Based Genetic Algorithm}

\author{Randy Lemons}
\affiliation{SLAC National Accelerator Laboratory and Stanford University, 2575 Sand Hill
Rd, Menlo Park, CA 94025, USA}
\affiliation{Colorado School of Mines, 1500 Illinois St, Golden, CO 80401, USA}
\email{rlemons@slac.stanford.edu}
\author{Sergio Carbajo}
\affiliation{SLAC National Accelerator Laboratory and Stanford University, 2575 Sand Hill
Rd, Menlo Park, CA 94025, USA}
\affiliation{Colorado School of Mines, 1500 Illinois St, Golden, CO 80401, USA}

\begin{abstract}

    
We present a numerical method for the reconstruction and optimization of complex field synthesis using coherent pulse combination systems. A genetic algorithm utilizing a Fourier optics based propagation method is developed for accurate convergence and modeling of near and far field distributions, achieving better than $\pi/10$ phase accuracy in reconstructed input parameters.


    
\end{abstract}

\maketitle

\section{Introduction}

The structuring of laser light beyond the simplistic regimes of conventional spatio-temporal distributions has taken hold in optical research within the past decades. New beams with non-diffracting vector distributions or optical vortices or ever increasing amounts of angular momentum are being synthesized and reported on, with each new structure finding use in the exploration of world around us. Non-diffracting beams are being exploited in nonlinear and biological imaging applications\cite{ND:mazilu2010light}, optical vortices have found use in very different fields such as optical trapping\cite{OV:ng2010theory} and micromachining\cite{OV:hamazaki2010optical}, and high-order momentum beams are enabling high throughput communications\cite{HM:willner2017recent} and particle manipulation\cite{HM:franke2017optical}.

Coherent combination has recently emerged for the synthesis of arbitrary intensity and phase profiles. By tuning the phase differences between adjacent beams, the combined field can be structured\cite{CC:r2020integrated}, directed\cite{CC:anderegg2006coherently}, and improved in the presence of propagation noise\cite{CC:zhou2009coherent}. However, the multi-element nature of these systems introduces a parameter space that is too large for it to be effectively investigated without computational aid. This is exacerbated as the number of elements grows, which is desirable for the creation of more complicated and refined fields. However, optimization techniques for this type of multi-dimensional problem are not scarce. With the advent of machine learning and computer vision, an innumerable list of optimization techniques has arisen. One such algorithm is the Genetic Algorithm (GA) based off the principles behind Darwin's theory of evolution.

GA's have been implemented with success before in optical applications for the tailoring of complex field structure. Ye et al. and Evans et al. have demonstrated the applicability GA for the design of optical elements\cite{GA:ye2001genetic} and shape optimization\cite{GA:evans1998design}. In this paper, a GA is implemented for the reconstruction and optimization of light structures arising from coherent combination systems. All propagation is modeled with the angular spectrum method which, under certain numerical constraints\cite{FP:buitrago2019non}, can be used to accurately model the evolution of light in the near and far fields\cite{FP:khareFourierOptics}. Finally we present two examples where the initial parameters are recovered based on near and far field intensity distributions.


\section{Methods}

\subsection{Angular Spectrum Fourier Propagation}

\subsubsection{The Wave Equation}






For the modeling of coherent combination systems we restrict ourselves to the case of free-space propagation without free-charges. As such the most general governing equation is the homogeneous wave-equation for electromagnetic waves

\begin{equation}
    \frac{1}{c^2} \frac{\partial^2 \mathbf{U}}{\partial t^2} - \nabla^2 \mathbf{U} = 0, \label{eqn:Wave_Simple}
\end{equation}
where $\mathbf{U}$ is used to represent either the electric or magnetic field. To further simplify evaluation of Eqn. \eqref{eqn:Wave_Simple} we can adopt the ansatz that $\mathrm{U}$ is separable into spatial and temporal parts such that $\mathrm{U} = u(\mathbf{r})v(t)$, where $\mathbf{r}$ represents a spatial vector. This decouples Eq. \eqref{eqn:Wave_Simple} into two independent equations
\begin{align}
        \left( \nabla^2 + k^2 \right) u(\mathbf{r}) &= 0, \label{eqn:Helm_Space} \\
        \left( \frac{\partial^2}{\partial t^2} + k^2 c^2 \right) v(t) &= 0, \label{eqn:Helm_Time}
\end{align}
where $k$ is the separation constant which we define as the wavevector by convention. We are concerned with the spatial distribution of light and will only focus on Eqn. \eqref{eqn:Helm_Space} moving forward. 
This spatial equation is known as the Helmholtz equation and is valid for monochromatic waves as long as the medium remains homogeneous, any diffracting apertures are much larger than the wavelength, and the divergence of light is much less than one radian.





\subsubsection{Angular Spectrum Solution}





The angular spectrum method (ASM) describes the propagation of waves that satisfy the Helmholtz equation. As such, all fields are described as an linear combination of directionally varying plane waves which are natural solutions to Eqn. \eqref{eqn:Helm_Space}. By convention, the light is assumed to be propagating arbitrary along positive $z$ and the field is evaluated in the transverse $xy$-plane at some location of constant $z$. Additionally the ASM utilizes the linear nature of optical systems to describe a propagated field as the convolution of an input field with the impulse response of the system such that
\begin{equation}
    E(x,y,z_2) = E(x,y,z_1) \ast H(x,y,z_2-z_1). \label{eqn:ASM_Begin}
\end{equation}

To begin finding solutions to Eqn. \eqref{eqn:ASM_Begin} we define the Fourier transform of the field in the $xy$-plane as
\begin{align}
    \begin{split}
        \hat{E}(f_x,f_y,z) &= \mathcal{F}\{E(x,y,z)\} \\
        &= \int \!\!\! \int_{-\infty}^{\infty} E(x,y,z) e^{-i 2 \pi (f_x x+f_y y)} dx dy,\label{eqn:Fourier} \\
    \end{split}
\end{align}
and the inverse transform as
\begin{align}
    \begin{split}
        E(x,y,z) &= \mathcal{F}^{-1}\{\hat{E}(f_x,f_y,z)\} \\
        &= \int \!\!\! \int_{-\infty}^{\infty} \hat{E}(f_x,f_y,z)  e^{i 2 \pi (f_x x+f_y y)} df_x df_y,\label{eqn:InvFourier}
    \end{split}
\end{align}
where $\mathbf{f_i}, \, i =\{x,y\}$ are the Fourier conjugate variables to position known as spatial frequencies comprising the angular spectrum and leading to the naming of this method. Upon plugging  Eqn. \eqref{eqn:InvFourier} into Eqn. \eqref{eqn:Helm_Space} we end up with 
\begin{equation*}
    \left[-2\pi \left(f_x^2+f_y^2\right)+\frac{\partial^2}{\partial z^2} + k^2\right]\hat{E}(f_x,f_y,z) = 0.
\end{equation*}
By defining $k_z = \sqrt{k^2-2\pi(f_x^2+f_y^2)}$ this equation simplifies to an alternative representation of the Helmholtz equation and has solutions of the form $\hat{E}(f_x,f_y,z) = C_1e^{i k_z z}$. Setting $z=0$, $C_1 = \hat{E}(f_x,f_y,0)$ and the angular spectrum of the field at some location $z$ is given by the angular spectrum of the input field modulated by the propagation of a wave in the $z$ direction with wavenumber $k_z$ and wavelength $\lambda$. Plugging this result back into Eqn. \eqref{eqn:InvFourier}, we get the result

\begin{multline}
    E(x,y,z) = \int \!\!\! \int_{-\infty}^{\infty} \hat{E}(f_x,f_y,0) e^{i k z \sqrt{1 - \lambda^2 \left( f_x^2 + f_y^2 \right)}} \\
    e^{i 2 \pi (f_x x+f_y y)} df_x df_y,\label{eqn:ASM_Full}
\end{multline}

Looking at Eqn. \eqref{eqn:ASM_Begin} we can use to properties of Fourier transforms to change the convolution to multiplication in the reciprocal space such that
\begin{multline}
    E(x,y,z_2) = \\
    \mathcal{F}^{-1} \{ \mathcal{F}\{E(x,y,z_1)\} \ast \mathcal{F}\{H(x,y,z_2-z_1)\}\}. \label{eqn:ASM_Four}
\end{multline}
Under this form it becomes evident that the factor $e^{i k z \sqrt{1 - \lambda^2 \left( f_x^2 + f_y^2 \right)}}$ in Eqn. \eqref{eqn:ASM_Full} is nothing more than the Fourier transform of the impulse response of free-space to a monochromatic wave\cite{FP:voelz2011computational}. In this case the Fourier transform of the impulse response is called the optical transfer function (OTF). Additionally any linear system can be modeled by the ASM given the OTF is known.


 

\subsection{Genetic Algorithm}
The GA is a global optimization technique where an optimal solution is found via informed stochastic search. This type of algorithm is efficient at finding solutions in large multi-dimensional parameters spaces found in such problems as the knapsack and traveling salesmen\cite{GA:braun1990solving}\cite{GA:chu1998genetic}. Mirroring Darwin's theory, a population of individual solutions which each contain genetic information corresponding to the variables of the problem at hand are initialized randomly within the domain. These possible individuals are then evaluated and ranked according to a fitness function as an analog for natural selection. Finally a new generation of solutions is created from this population by preferentially selecting optimal individuals as parents and mixing their genes.

\begin{figure}[ht!]
	\centering
	\includegraphics[width=1\linewidth]{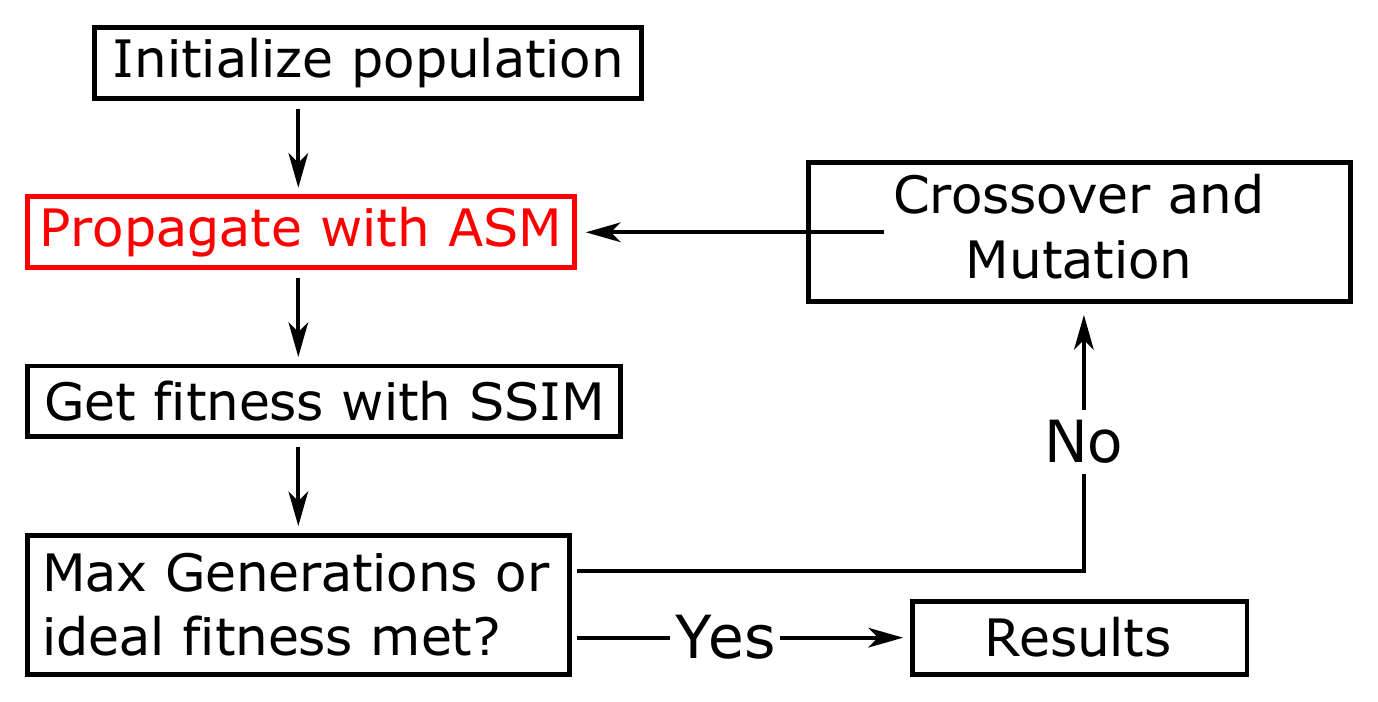}
	\caption{Block diagram of our GA with the addition of Fourier propagation in red}
  \label{fig:GA_Block}
\end{figure}

For our implementation each individual in the population is a single field which can be propagated with the ASM where the genes are the initial field parameters. The entire population is initialized at random values throughout the parameter space, propagated to the plane under investigation and then compared against either simulated or experimental data. In order to compare the GA population to real data we needed a fitness function that worked on an easily accessible observable such as camera intensity images. For this reason we settled on a version of the structural similarity index (SSIM) which is a normalized metric comparing the local intensity, structure, and contrast between two images. The modified version of the SSIM below is a true distance metric detailed by Brunet et al\cite{SSIM:brunet2011mathematical}. 
\begin{equation}
    D = \sqrt{2 - \frac{2 \mu_{12}+c_1}{\mu_1^2+\mu_2^2+c_1} - \frac{2 S_{12}+c_2}{S_1^2+S_2^2+c_1} }, \label{eqn:SSIM}
\end{equation}
where $\mu_{i}$ refers to the local mean and $s_{i}$ the local variance and $c_i$ is a small constants added for numerical stability as the two images become more similar. This formula results in a matrix the size of the input images and the normalized SSIM metric is the mean of Eqn. \eqref{eqn:SSIM}.
For crossover and recreation we take the top 10\% and a randomly selected subset of the remaining 90\% such that a total of 15\% of the population survives. The next generation is then created by selected two different parents and randomly mixing half of the genetic information from each to create a new individual. Mutation where a small percentage of the total genes in the pool are reinitialized to new positions throughout the parameter space is the last step before restarting the process at propagation.

\section{Results}

In order to test the GA we used experimental images from a seven beam, free space, coherent combination system developed for structured field synthesis rather than maximal intensity. The amplitude, phase offset, and whether the beam is on or off can be tuned for each individual beam in the array as described in Lemons et al.\cite{CC:r2020integrated}. The seven beams are manipulated in single mode fiber, placed in a hexagonal tiled array and coupled out to free space via a micro lens array (MLA). As such, each beam has an unknown but similar curvature as well as an aperture on the Gaussian profiles induced by the MLA. In order to characterize the effectiveness of the GA on recreating near field parameters based on propagated images we present two reconstruction scenarios: a single beam in the near field, and the seven beam array with induced phase offsets in the far field. The parameters recovered from the single beam in the near field are used to constrain the size of the aperture the curvature for the far field reconstruction.

\begin{figure}[ht!]
	\centering
	\includegraphics[width=1\linewidth]{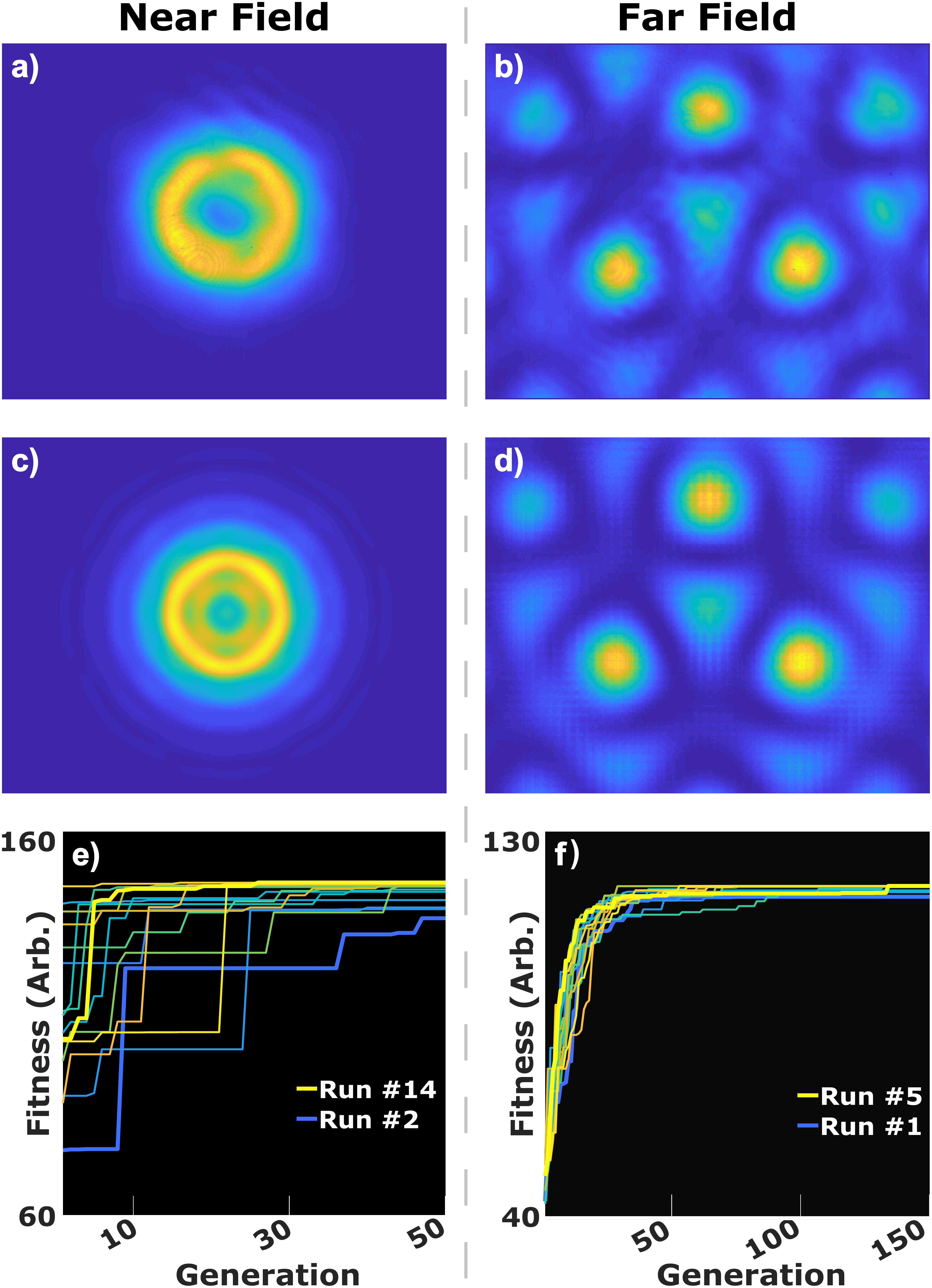}
	\caption{Results from the GA running on the outer six beams in the far field. The reconstructed intensity profile (b) qualitatively matches the experimental image (a) well with recreating the trefoil-like distribution. The fitness for all generations and runs in shown in c.}
  \label{fig:Near_Far_Fig}
\end{figure}

In the near field the seven beams have a donut-like intensity profile (Fig. \ref{fig:Near_Far_Fig}a). To reconstruct these parameters all but the center beam in the array is turned off to eliminate interference from the nearby beams. The camera was then placed 1 m away from the MLA, centered on the remaining beam. The GA was initialized with 50 individuals having free parameters of Gaussian waist size and the curvature added by the MLA with a total of $5.4 \cdot 10^7$ combinations. It was then run for 100 generations over 15 different runs in order to demonstrate consistent convergence.

For all 15 runs, the fitness of the best individual per generation (Fig. \ref{fig:Near_Far_Fig}e) converges to nearly identical values. Some runs also started with highly fit individuals due to the relatively small parameter space. The large convergence at the beginning is due to SSIM increasing quickly for large structures. The small increases from there on out are characterized by the GA refining the values closer to the true value. Across the 15 runs the average curvature was -1.64 $\pm .07$ m corresponding to a slightly converging beam with an average waist of 3.2 $\pm .8$ mm. The MLA in experiment has a hard aperture of 3 mm on each beam and is designed to collimate light with high numerical aperture outcoupled from optical fibers. As such the divergence from collimated and slight over filling of each aperture are reasonable to expect.

For the far field test the center beam in the array was turned off and the outside six beams were given an alternating phase offset of zero and $\pi/2$ such that the difference between any two adjacent beams was $\pm \pi/2$. The image was taken from the first frame of the camera after the induced shift to ensure the captured profile most closely matched the induced shift. The far field combination of this arrangement, shown in Fig. \ref{fig:Near_Far_Fig}b, is characterized by a distribution reminiscent of the international radiation symbol. This time the GA contained 120 individuals over 150 generations (still with 15 runs) and was allowed to change the phase offset and amplitude of each beam for $3.62 \cdot 10^{18}$ permutations. Each individual was also initialized with -1.65 m curvature and 3 mm waist size based on the near field test.

\begin{figure}[ht!]
	\centering
	\includegraphics[width=1\linewidth]{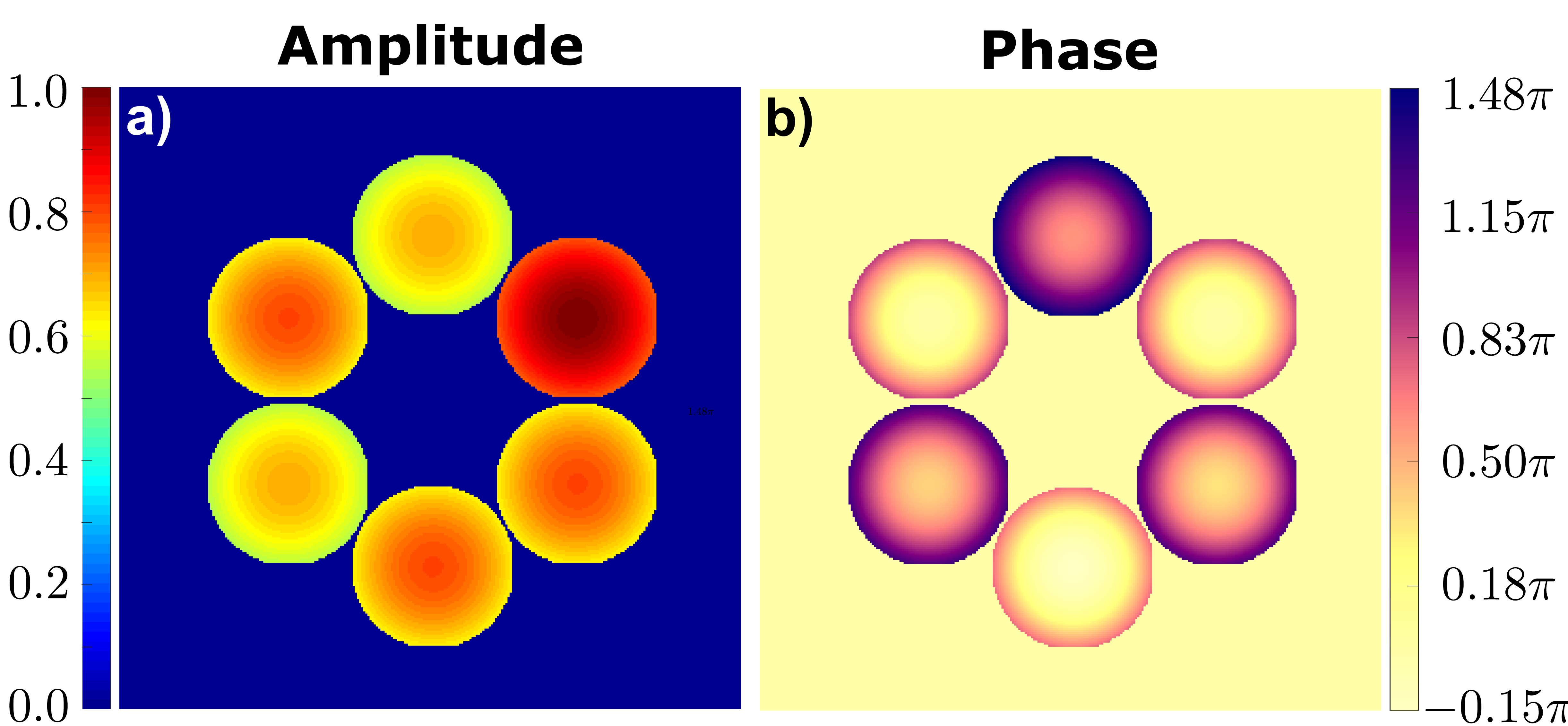}
	\caption{Initial parameters of the best individual across all runs in the far field test. The amplitude of the beams is shown in a) while the stair step phase profile between any two adjacent beams and curvature of the wavefronts is shown in b).}
  \label{fig:Far_Vals}
\end{figure}

Again the fitness (Fig. \ref{fig:Near_Far_Fig}f) converges to nearly identical values toward the end of each run with large convergence at the beginning of each run and small increases from there on out. For this test, no run started with a high fitness individual due to the the significantly larger parameter space displaying the true benefit of the GA to a problem like this. In this case each of the six beams converged to slightly differing values close to zero or $\pi/2$ seen in Fig. \ref{fig:Far_Vals}b with slightly differing amplitude (Fig. \ref{fig:Far_Vals}a).

 Slight amplitude differences in the beams are expected due to different losses for each beam before combination. The small deviations from strict $\pi/2$ stair stepping likely arise from the time difference between adjustment and image capture. For any given beam the deviation between reconstructed phase values across all runs was less than $0.09 \pi$ and around than $10\%$ for amplitude.

\section{Conclusion}

 We have presented a genetic algorithm that makes use of modern image processing techniques and proven propagation techniques to recover and explore the vast space opened up from coherent combination systems with arbitrary near field parameters. We successfully recovered unknown near field parameters and used them to reconstruct sub-wavelength phase differences between beams based on far field intensity distributions. Though this algorithm is able to effectively work in both the near and field regimes, improving the propagation technique to include modern anti-aliasing techniques will increase the valid distances. This would enable the modeling and optimization of the systems used in directed energy and laser propulsion applications.
 Additionally the accuracy of the GA can be improved by tailoring of the fitness function to individual setups and convergence time by implementation of more sophisticated crossover and mutation techniques. Increased accuracy of the optimized fields could be exploited for the generation of increasing exotic modes used in optical communication multiplexing in fiber.


\bibliographystyle{aipauth4-2}
\bibliography{bib}

\begin{thebibliography}{16}%
\makeatletter
\providecommand \@ifxundefined [1]{%
 \@ifx{#1\undefined}
}%
\providecommand \@ifnum [1]{%
 \ifnum #1\expandafter \@firstoftwo
 \else \expandafter \@secondoftwo
 \fi
}%
\providecommand \@ifx [1]{%
 \ifx #1\expandafter \@firstoftwo
 \else \expandafter \@secondoftwo
 \fi
}%
\providecommand \natexlab [1]{#1}%
\providecommand \enquote  [1]{``#1''}%
\providecommand \bibnamefont  [1]{#1}%
\providecommand \bibfnamefont [1]{#1}%
\providecommand \citenamefont [1]{#1}%
\providecommand \href@noop [0]{\@secondoftwo}%
\providecommand \href [0]{\begingroup \@sanitize@url \@href}%
\providecommand \@href[1]{\@@startlink{#1}\@@href}%
\providecommand \@@href[1]{\endgroup#1\@@endlink}%
\providecommand \@sanitize@url [0]{\catcode `\\12\catcode `\$12\catcode
  `\&12\catcode `\#12\catcode `\^12\catcode `\_12\catcode `\%12\relax}%
\providecommand \@@startlink[1]{}%
\providecommand \@@endlink[0]{}%
\providecommand \url  [0]{\begingroup\@sanitize@url \@url }%
\providecommand \@url [1]{\endgroup\@href {#1}{\urlprefix }}%
\providecommand \urlprefix  [0]{URL }%
\providecommand \Eprint [0]{\href }%
\providecommand \doibase [0]{https://doi.org/}%
\providecommand \selectlanguage [0]{\@gobble}%
\providecommand \bibinfo  [0]{\@secondoftwo}%
\providecommand \bibfield  [0]{\@secondoftwo}%
\providecommand \translation [1]{[#1]}%
\providecommand \BibitemOpen [0]{}%
\providecommand \bibitemStop [0]{}%
\providecommand \bibitemNoStop [0]{.\EOS\space}%
\providecommand \EOS [0]{\spacefactor3000\relax}%
\providecommand \BibitemShut  [1]{\csname bibitem#1\endcsname}%
\let\auto@bib@innerbib\@empty
\bibitem [{\citenamefont {Anderegg}\ \emph {et~al.}(2006)\citenamefont
  {Anderegg}, \citenamefont {Brosnan}, \citenamefont {Cheung}, \citenamefont
  {Epp}, \citenamefont {Hammons}, \citenamefont {Komine}, \citenamefont
  {Weber},\ and\ \citenamefont {Wickham}}]{CC:anderegg2006coherently}%
  \BibitemOpen
  \bibfield  {author} {\bibinfo {author} {\bibnamefont {Anderegg},
  \bibfnamefont {J.}}, \bibinfo {author} {\bibnamefont {Brosnan}, \bibfnamefont
  {S.}}, \bibinfo {author} {\bibnamefont {Cheung}, \bibfnamefont {E.}},
  \bibinfo {author} {\bibnamefont {Epp}, \bibfnamefont {P.}}, \bibinfo {author}
  {\bibnamefont {Hammons}, \bibfnamefont {D.}}, \bibinfo {author} {\bibnamefont
  {Komine}, \bibfnamefont {H.}}, \bibinfo {author} {\bibnamefont {Weber},
  \bibfnamefont {M.}}, and\ \bibinfo {author} {\bibnamefont {Wickham},
  \bibfnamefont {M.}},\ }in\ \href@noop {} {\emph {\bibinfo {booktitle} {Fiber
  Lasers III: Technology, Systems, and Applications}}},\ Vol.\ \bibinfo
  {volume} {6102}\ (\bibinfo {organization} {International Society for Optics
  and Photonics},\ \bibinfo {year} {2006})\ p.\ \bibinfo {pages}
  {61020U}\BibitemShut {NoStop}%
\bibitem [{\citenamefont {Braun}(1990)}]{GA:braun1990solving}%
  \BibitemOpen
  \bibfield  {author} {\bibinfo {author} {\bibnamefont {Braun}, \bibfnamefont
  {H.}},\ }in\ \href@noop {} {\emph {\bibinfo {booktitle} {International
  Conference on Parallel Problem Solving from Nature}}}\ (\bibinfo
  {organization} {Springer},\ \bibinfo {year} {1990})\ pp.\ \bibinfo {pages}
  {129--133}\BibitemShut {NoStop}%
\bibitem [{\citenamefont {Brunet}, \citenamefont {Vrscay},\ and\ \citenamefont
  {Wang}(2011)}]{SSIM:brunet2011mathematical}%
  \BibitemOpen
  \bibfield  {author} {\bibinfo {author} {\bibnamefont {Brunet}, \bibfnamefont
  {D.}}, \bibinfo {author} {\bibnamefont {Vrscay}, \bibfnamefont {E.~R.}}, and\
  \bibinfo {author} {\bibnamefont {Wang}, \bibfnamefont {Z.}},\ }\href@noop {}
  {\bibfield  {journal} {\bibinfo  {journal} {IEEE Transactions on Image
  Processing}\ }\textbf {\bibinfo {volume} {21}},\ \bibinfo {pages} {1488}
  (\bibinfo {year} {2011})}\BibitemShut {NoStop}%
\bibitem [{\citenamefont {Buitrago-Duque}\ and\ \citenamefont
  {Garcia-Sucerquia}(2019)}]{FP:buitrago2019non}%
  \BibitemOpen
  \bibfield  {author} {\bibinfo {author} {\bibnamefont {Buitrago-Duque},
  \bibfnamefont {C.}}and\ \bibinfo {author} {\bibnamefont {Garcia-Sucerquia},
  \bibfnamefont {J.}},\ }\href@noop {} {\bibfield  {journal} {\bibinfo
  {journal} {Applied Optics}\ }\textbf {\bibinfo {volume} {58}},\ \bibinfo
  {pages} {G11} (\bibinfo {year} {2019})}\BibitemShut {NoStop}%
\bibitem [{\citenamefont {Chu}\ and\ \citenamefont
  {Beasley}(1998)}]{GA:chu1998genetic}%
  \BibitemOpen
  \bibfield  {author} {\bibinfo {author} {\bibnamefont {Chu}, \bibfnamefont
  {P.~C.}}and\ \bibinfo {author} {\bibnamefont {Beasley}, \bibfnamefont
  {J.~E.}},\ }\href@noop {} {\bibfield  {journal} {\bibinfo  {journal} {Journal
  of heuristics}\ }\textbf {\bibinfo {volume} {4}},\ \bibinfo {pages} {63}
  (\bibinfo {year} {1998})}\BibitemShut {NoStop}%
\bibitem [{\citenamefont {Evans}\ and\ \citenamefont
  {Shealy}(1998)}]{GA:evans1998design}%
  \BibitemOpen
  \bibfield  {author} {\bibinfo {author} {\bibnamefont {Evans}, \bibfnamefont
  {N.~C.}}and\ \bibinfo {author} {\bibnamefont {Shealy}, \bibfnamefont
  {D.~L.}},\ }\href@noop {} {\bibfield  {journal} {\bibinfo  {journal} {Applied
  Optics}\ }\textbf {\bibinfo {volume} {37}},\ \bibinfo {pages} {5216}
  (\bibinfo {year} {1998})}\BibitemShut {NoStop}%
\bibitem [{\citenamefont {Franke-Arnold}(2017)}]{HM:franke2017optical}%
  \BibitemOpen
  \bibfield  {author} {\bibinfo {author} {\bibnamefont {Franke-Arnold},
  \bibfnamefont {S.}},\ }\href@noop {} {\bibfield  {journal} {\bibinfo
  {journal} {Philosophical Transactions of the Royal Society A: Mathematical,
  Physical and Engineering Sciences}\ }\textbf {\bibinfo {volume} {375}},\
  \bibinfo {pages} {20150435} (\bibinfo {year} {2017})}\BibitemShut {NoStop}%
\bibitem [{\citenamefont {Hamazaki}\ \emph {et~al.}(2010)\citenamefont
  {Hamazaki}, \citenamefont {Morita}, \citenamefont {Chujo}, \citenamefont
  {Kobayashi}, \citenamefont {Tanda},\ and\ \citenamefont
  {Omatsu}}]{OV:hamazaki2010optical}%
  \BibitemOpen
  \bibfield  {author} {\bibinfo {author} {\bibnamefont {Hamazaki},
  \bibfnamefont {J.}}, \bibinfo {author} {\bibnamefont {Morita}, \bibfnamefont
  {R.}}, \bibinfo {author} {\bibnamefont {Chujo}, \bibfnamefont {K.}}, \bibinfo
  {author} {\bibnamefont {Kobayashi}, \bibfnamefont {Y.}}, \bibinfo {author}
  {\bibnamefont {Tanda}, \bibfnamefont {S.}}, and\ \bibinfo {author}
  {\bibnamefont {Omatsu}, \bibfnamefont {T.}},\ }\href@noop {} {\bibfield
  {journal} {\bibinfo  {journal} {Optics express}\ }\textbf {\bibinfo {volume}
  {18}},\ \bibinfo {pages} {2144} (\bibinfo {year} {2010})}\BibitemShut
  {NoStop}%
\bibitem [{\citenamefont {Khare}(2015)}]{FP:khareFourierOptics}%
  \BibitemOpen
  \bibfield  {author} {\bibinfo {author} {\bibnamefont {Khare}, \bibfnamefont
  {K.}},\ }\href@noop {} {\emph {\bibinfo {title} {Fourier optics and
  computational imaging}}},\ Ane/Athena Bks\ (\bibinfo  {publisher} {John Wiley
  and Sons, Limited},\ \bibinfo {address} {Chichester, West Sussex, England},\
  \bibinfo {year} {2015})\BibitemShut {NoStop}%
\bibitem [{\citenamefont {Lemons}\ \emph {et~al.}(2020)\citenamefont {Lemons},
  \citenamefont {Liu}, \citenamefont {Frisch}, \citenamefont {Fry},
  \citenamefont {Robinson}, \citenamefont {Smith},\ and\ \citenamefont
  {Carbajo}}]{CC:r2020integrated}%
  \BibitemOpen
  \bibfield  {author} {\bibinfo {author} {\bibnamefont {Lemons}, \bibfnamefont
  {R.}}, \bibinfo {author} {\bibnamefont {Liu}, \bibfnamefont {W.}}, \bibinfo
  {author} {\bibnamefont {Frisch}, \bibfnamefont {J.~C.}}, \bibinfo {author}
  {\bibnamefont {Fry}, \bibfnamefont {A.}}, \bibinfo {author} {\bibnamefont
  {Robinson}, \bibfnamefont {J.}}, \bibinfo {author} {\bibnamefont {Smith},
  \bibfnamefont {S.}}, and\ \bibinfo {author} {\bibnamefont {Carbajo},
  \bibfnamefont {S.}},\ }\href@noop {} {\enquote {\bibinfo {title} {Integrated
  structured light architectures},}\ } (\bibinfo {year} {2020}),\ \Eprint
  {https://arxiv.org/abs/2003.14400} {arXiv:2003.14400 [physics.optics]}
  \BibitemShut {NoStop}%
\bibitem [{\citenamefont {Mazilu}\ \emph {et~al.}(2010)\citenamefont {Mazilu},
  \citenamefont {Stevenson}, \citenamefont {Gunn-Moore},\ and\ \citenamefont
  {Dholakia}}]{ND:mazilu2010light}%
  \BibitemOpen
  \bibfield  {author} {\bibinfo {author} {\bibnamefont {Mazilu}, \bibfnamefont
  {M.}}, \bibinfo {author} {\bibnamefont {Stevenson}, \bibfnamefont {D.~J.}},
  \bibinfo {author} {\bibnamefont {Gunn-Moore}, \bibfnamefont {F.}}, and\
  \bibinfo {author} {\bibnamefont {Dholakia}, \bibfnamefont {K.}},\ }\href@noop
  {} {\bibfield  {journal} {\bibinfo  {journal} {Laser \& Photonics Reviews}\
  }\textbf {\bibinfo {volume} {4}},\ \bibinfo {pages} {529} (\bibinfo {year}
  {2010})}\BibitemShut {NoStop}%
\bibitem [{\citenamefont {Ng}, \citenamefont {Lin},\ and\ \citenamefont
  {Chan}(2010)}]{OV:ng2010theory}%
  \BibitemOpen
  \bibfield  {author} {\bibinfo {author} {\bibnamefont {Ng}, \bibfnamefont
  {J.}}, \bibinfo {author} {\bibnamefont {Lin}, \bibfnamefont {Z.}}, and\
  \bibinfo {author} {\bibnamefont {Chan}, \bibfnamefont {C.}},\ }\href@noop {}
  {\bibfield  {journal} {\bibinfo  {journal} {Physical review letters}\
  }\textbf {\bibinfo {volume} {104}},\ \bibinfo {pages} {103601} (\bibinfo
  {year} {2010})}\BibitemShut {NoStop}%
\bibitem [{\citenamefont {Voelz}(2011)}]{FP:voelz2011computational}%
  \BibitemOpen
  \bibfield  {author} {\bibinfo {author} {\bibnamefont {Voelz}, \bibfnamefont
  {D.}}\ }(\bibinfo {organization} {Society of Photo-Optical Instrumentation
  Engineers},\ \bibinfo {year} {2011})\BibitemShut {NoStop}%
\bibitem [{\citenamefont {Willner}\ \emph {et~al.}(2017)\citenamefont
  {Willner}, \citenamefont {Ren}, \citenamefont {Xie}, \citenamefont {Yan},
  \citenamefont {Li}, \citenamefont {Zhao}, \citenamefont {Wang}, \citenamefont
  {Tur}, \citenamefont {Molisch},\ and\ \citenamefont
  {Ashrafi}}]{HM:willner2017recent}%
  \BibitemOpen
  \bibfield  {author} {\bibinfo {author} {\bibnamefont {Willner}, \bibfnamefont
  {A.~E.}}, \bibinfo {author} {\bibnamefont {Ren}, \bibfnamefont {Y.}},
  \bibinfo {author} {\bibnamefont {Xie}, \bibfnamefont {G.}}, \bibinfo {author}
  {\bibnamefont {Yan}, \bibfnamefont {Y.}}, \bibinfo {author} {\bibnamefont
  {Li}, \bibfnamefont {L.}}, \bibinfo {author} {\bibnamefont {Zhao},
  \bibfnamefont {Z.}}, \bibinfo {author} {\bibnamefont {Wang}, \bibfnamefont
  {J.}}, \bibinfo {author} {\bibnamefont {Tur}, \bibfnamefont {M.}}, \bibinfo
  {author} {\bibnamefont {Molisch}, \bibfnamefont {A.~F.}}, and\ \bibinfo
  {author} {\bibnamefont {Ashrafi}, \bibfnamefont {S.}},\ }\href@noop {}
  {\bibfield  {journal} {\bibinfo  {journal} {Philosophical Transactions of the
  Royal Society A: Mathematical, Physical and Engineering Sciences}\ }\textbf
  {\bibinfo {volume} {375}},\ \bibinfo {pages} {20150439} (\bibinfo {year}
  {2017})}\BibitemShut {NoStop}%
\bibitem [{\citenamefont {Ye}, \citenamefont {Yuan},\ and\ \citenamefont
  {Zhou}(2001)}]{GA:ye2001genetic}%
  \BibitemOpen
  \bibfield  {author} {\bibinfo {author} {\bibnamefont {Ye}, \bibfnamefont
  {J.}}, \bibinfo {author} {\bibnamefont {Yuan}, \bibfnamefont {X.}}, and\
  \bibinfo {author} {\bibnamefont {Zhou}, \bibfnamefont {G.}},\ }in\ \href@noop
  {} {\emph {\bibinfo {booktitle} {Design, Fabrication, and Characterization of
  Photonic Devices II}}},\ Vol.\ \bibinfo {volume} {4594}\ (\bibinfo
  {organization} {International Society for Optics and Photonics},\ \bibinfo
  {year} {2001})\ pp.\ \bibinfo {pages} {118--127}\BibitemShut {NoStop}%
\bibitem [{\citenamefont {Zhou}\ \emph {et~al.}(2009)\citenamefont {Zhou},
  \citenamefont {Liu}, \citenamefont {Wang}, \citenamefont {Ma}, \citenamefont
  {Ma}, \citenamefont {Xu},\ and\ \citenamefont {Guo}}]{CC:zhou2009coherent}%
  \BibitemOpen
  \bibfield  {author} {\bibinfo {author} {\bibnamefont {Zhou}, \bibfnamefont
  {P.}}, \bibinfo {author} {\bibnamefont {Liu}, \bibfnamefont {Z.}}, \bibinfo
  {author} {\bibnamefont {Wang}, \bibfnamefont {X.}}, \bibinfo {author}
  {\bibnamefont {Ma}, \bibfnamefont {Y.}}, \bibinfo {author} {\bibnamefont
  {Ma}, \bibfnamefont {H.}}, \bibinfo {author} {\bibnamefont {Xu},
  \bibfnamefont {X.}}, and\ \bibinfo {author} {\bibnamefont {Guo},
  \bibfnamefont {S.}},\ }\href@noop {} {\bibfield  {journal} {\bibinfo
  {journal} {IEEE journal of selected topics in quantum electronics}\ }\textbf
  {\bibinfo {volume} {15}},\ \bibinfo {pages} {248} (\bibinfo {year}
  {2009})}\BibitemShut {NoStop}%
\end{thebibliography}%

\end{document}